\documentclass[%
 reprint, amsmath,amssymb,
 aps,
]{revtex4-2}

\usepackage{graphicx}
\usepackage{dcolumn}
\usepackage{bm}
\usepackage{siunitx}
\usepackage{braket}
\usepackage{lmodern}

\begin{document}

\preprint{APS/123-QED}

\title{Diffusive-like Redistribution in State-changing Collisions between Rydberg Atoms and Ground State Atoms}
\author{Philipp Geppert}
\author{Max Althön}
\author{Daniel Fichtner}
\author{Herwig Ott}
\email{ott@physik.uni-kl.de}
\affiliation{%
 Department of Physics and Research Center OPTIMAS, Technische Universität Kaiserslautern, 67663 Kaiserslautern, Germany
}%

\date{\today}
             
\begin{abstract}
We report on the study of state-changing collisions between Rydberg atoms and ground state atoms. We employ high-resolution momentum spectroscopy to identify the final states. In contrast to previous studies, we find that the outcome of such collisions is not limited to a single hydrogenic manifold. We observe a redistribution of population over a wide range of final states. We also find that even the decay to states with the same $l$ as the initial state, but different principal quantum number $n$ is possible. We model the underlying physical process in the framework of a short-lived Rydberg quasi-molecular complex, where a charge exchange process gives rise to an oscillating electric field that causes transitions within the Rydberg manifold. The distribution of final states shows a diffusive-like behavior.  
\end{abstract}

\maketitle

\section{\label{sec:introduction}Introduction}
The understanding of collisions between Rydberg atoms and ground state atoms has a long history and dates back to seminal work done by Enrico Fermi \cite{Fermi1934,Fermi2008}. Today, such processes are important for low-temperature plasma physics \cite{Rolston2014}, astrophysical plasmas \cite{Klyucharev2010} and in ultracold atom experiments, which have found in Rydberg physics a perfect match to explore ultracold chemistry and many-body physics: on the one hand, the high control over the internal and external degrees of freedom in an ultracold atomic gas enables the study of new phenomena in the field of Rydberg physics, such as Rydberg molecules \cite{Shaffer2018}, Rydberg blockade \cite{Urban2009}, Rydberg antiblockade \cite{Weber2015,Amthor2010} and coherent many-body dynamics \cite{Barredo2015}. On the other hand, the same control can now be used to study established processes in a detailed fashion, thus unraveling the underlying microscopic physical mechanisms. This way, the state-resolved study of inelastic collisions and molecular decay processes involving Rydberg atoms has become possible.
 
Collisions between a Rydberg atom and a ground state atom can have several possible outcomes. Here, we are interested in the dissociation of both partners, where the Rydberg atom undergoes a transition to a lower-lying state, while the excess energy is converted into kinetic energy of both collision partners. Such collisions have been studied in detail by Schlagm\"uller \textit{et al.} \cite{Schlagmueller2016}. They are important for the understanding of recombination processes in plasmas, for the quantitative understanding of inelastic processes in Rydberg gases \cite{Porto2016} and the decay dynamics of ultralong-range Rydberg molecules.

The microscopic details of such a collision involve the physics of ultralong-range Rydberg molecules \cite{Shaffer2018}, where s- and p-wave scattering between the Rydberg electron and the ground state atom determine the potential energy landscape at large internuclear distances. At short internuclear distances, however, the covalent molecular binding mechanisms take over and dominate the molecular dynamics. The total scattering process therefore probes the potential energy landscape at all internuclear distances. Thus, the understanding of such a process requires the modeling of both, the ultralong-range potential energy landscape as well as that at short internuclear distances.

An experimental in-depth study requires the state-selective detection of the reaction products. Only then, it is possible to access branching ratios and selection rules and one can compare the experimental outcome to effective theoretical models. Magneto-Optical Trap Recoil Ion Momentum Spectroscopy (MOTRIMS)\cite{Wolf2000,Poel2001,Turkstra2001,Flechard2001,Nguyen2004,Blieck2008,Depaola2008,Schuricke2011,Fischer2012,Goetz2012,Hubele2015,Li2019} is such a technique, which has been used to perform momentum spectroscopy of atomic and molecular processes with high resolution. Inspired by the MOTRIMS technique, we have developed a new high resolution momentum microscope, which enables the study of state-resolved inelastic processes involving Rydberg atoms. We use this technique to measure the dissociation of ultralong-range rubidium Rydberg molecules for principal quantum numbers between $n=20$ and $n=27$.

Our manuscript is organized as follows. In section \ref{sectionII}, we introduce the dissociation process we are interested in and summarize previous results. The momentum microscope is described in section \ref{sectionIII} together with a layout of the experimental sequence. The results are discussed in section \ref{sectionIV}, while in section \ref{sectionV} we give an interpretation of our findings in terms of a diffusive-like redistribution between the Rydberg states during the collision.

\section{\label{sectionII}State-Changing Collisions within Rydberg molecules}
\begin{figure}[t]
	\includegraphics[width=\columnwidth]{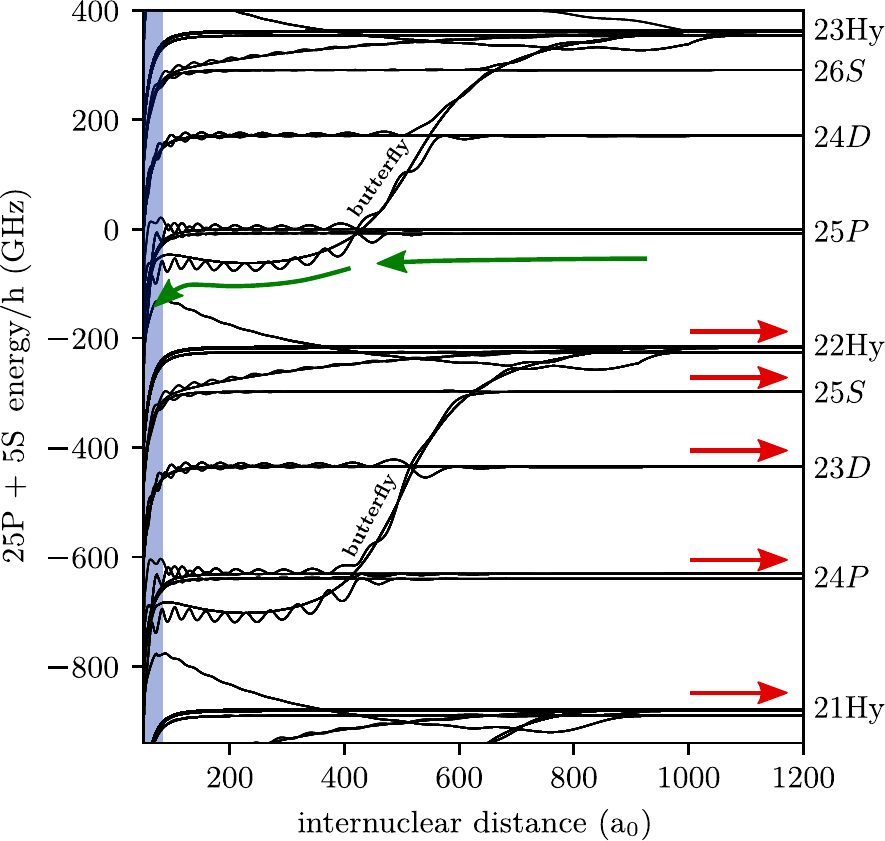}
	\caption{\label{fig:potential_energy_landscape} Adiabatic potential energy curves (PECs) of rubidium ultralong-range Rydberg molecules (ULRMs) around the 25$P$ state (see Appendix for details of the calculation). The annotations to the right denote the terms of the asymptotic free Rydberg states. Starting point of our studies is the preparation of rubidium ULRMs, which are bound vibrational states supported by the outermost potential wells at internuclear distances of up to $\SI{900}{a_0}$. As time evolves, the ground state atom tunnels towards the ionic core (green arrows), following the $R^{-4}$-interaction dominated PECs (blue shaded area) up to the region, where short-range molecular couplings are dominant. The red arrows indicate possible outcomes of a state-changing collision, where the release energy is translated into kinetic energy resulting in a dissociation of the molecule.}
\end{figure}
\begin{figure*}[t]
	\includegraphics[width=\textwidth]{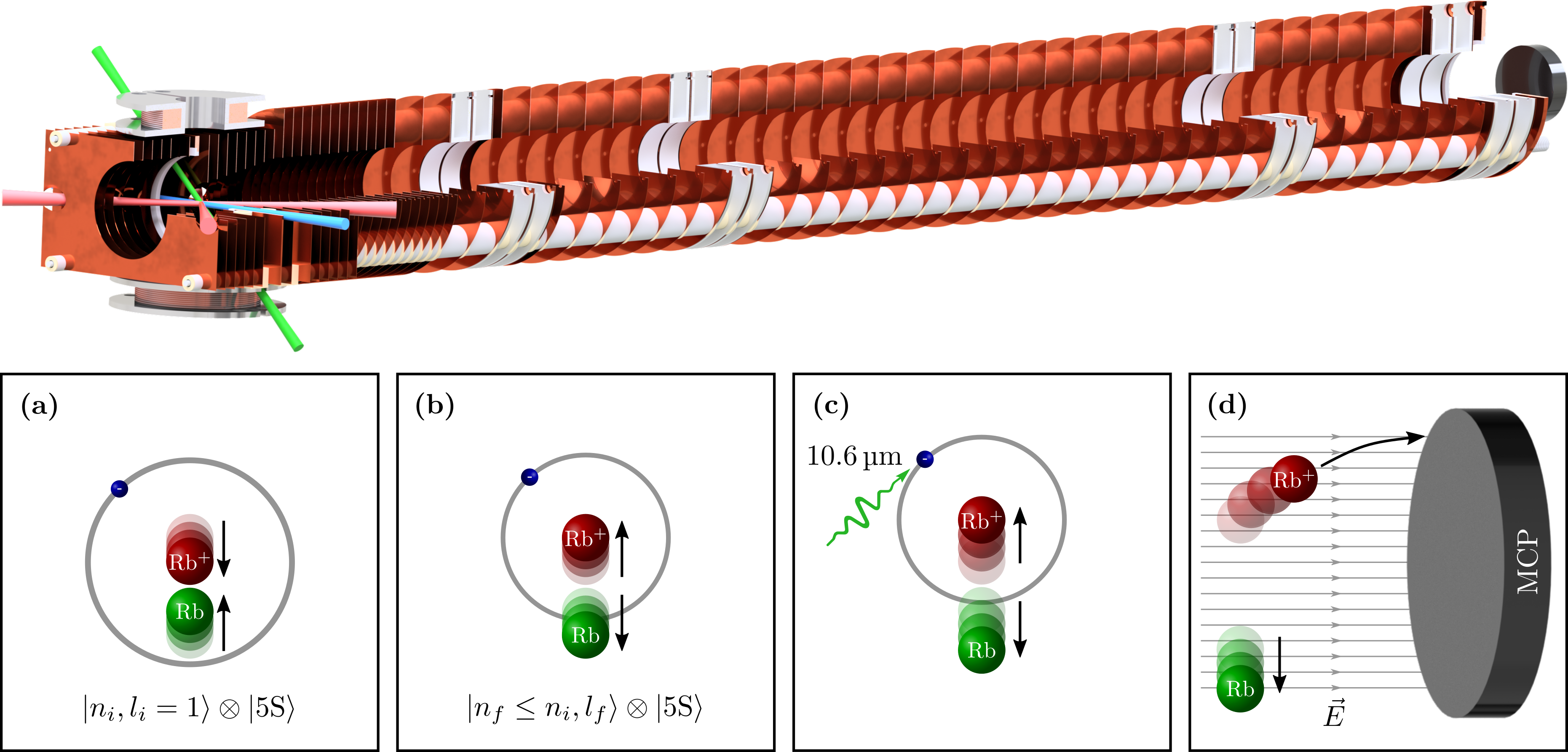}
	\caption{\label{fig:spectrometer} \textbf{Top:} CAD drawing of the experimental setup in three-quarter section view. Laser-cooled $^{87}$Rb atoms are trapped in a crossed optical dipole trap (red beams) and excited by a three photon transition (depicted as blue beam) to a Rydberg state. By using a high-power CO$_2$ laser pulse (green beam), the Rydberg atoms are photoionized efficiently. The ions then follow two subsequent homogeneous electric fields and traverse a field-free drift tube before hitting the position- and time-sensitive detector. This method allows the measurement of momentum distributions of initially neutral particles with high resolution. \textbf{Bottom:} Sketch, illustrating the state-changing collision process and the experimental procedure to measure the resulting momenta. (a) The experiment starts with the photoassociation of an ULRM, where the initial state of the Rydberg electron (blue) is $\ket{n_{\text{i}},l_{\text{i}}=1}$. Subsequently, the Rydberg core (red) and the neutral atom (green) approach each other. (b) Within the inelastic collision, the Rydberg electron changes its state (final principal quantum number $n_{\text{f}}$ is less than or equal to $n_{\text{i}}$). The final angular momentum quantum number $l_{\text{f}}$ may be any value between 0 and ($n_{\text{f}}$-1). The release energy is apportioned equally between the Rydberg atom and the perturber atom, which fly into opposite directions due to the momentum conservation. (c) The Rydberg atom is photoionized by a CO$_2$ laser pulse. (d) Electric fields guide the ionic core towards the detector without changing the transverse momentum. From the point of impact the momentum of the Rydberg atom can be inferred.}
\end{figure*}  

The interaction of a ground state atom and a Rydberg atom at large internuclear distances is mediated by low-energy scattering between the Rydberg electron and the ground state atom, also denoted as perturber atom. For rubidium, the potential energy at separations larger than the extent of the electronic wave function is given by the energy levels of an isolated Rydberg atom, $E_{nl} \propto -1/(n-\delta_l)^2$, where $n$ is the principal quantum number. The quantum defect $\delta_l$ causes a significant splitting of the potential energies only for angular momenta $l\leq2$. One therefore has to distinguish between energetically split low-$l$ ($S,P,D$) states with significant quantum defects and high-$l$ hydrogenic manifolds. For smaller internuclear distances, the scattering interaction between the Ryd\-berg electron and the ground state atom leads to oscillatory potentials, which support vibrational states. The resulting potential energy landscape is shown in Fig.\,\ref{fig:potential_energy_landscape}. We employ these so-called ultralong-range Ryd\-berg molecules (ULRMs) with bond lengths of $\approx \SI{900}{a_0}$ as starting point for our measurements and restrict ourselves to the regime of low principal quantum numbers owing to the $n^{-6}$ scaling of the outer wells' depths \cite{Fey2020}. As we use a three-photon transition to excite the Rydberg state, we specifically chose Rydberg n$P$ states as initial state. For a brief introduction to ULRMs, see Appendix A.

The ULRMs, which we excite, have a finite probability to tunnel towards the ionic core. This way, we mimic the dynamics of an inelastic collision between a free ground state atom and a Rydberg atom -- up to a small mismatch in energy, which stems from the binding energy of the Rydberg molecule. 

The molecular dynamic is initially dominated by the low-energy scattering between the electron and the ground state atom $ V_{p}(|\bm{r}-\bm{R}|)$ (second term in equation \ref{eq:pseudopotential}) and the ion-neutral polarization potential $V_{\text{c,g}}(R) \propto  R^{-4}$ (\ref{eq:ion-neutral-potential}). In case of alkali atoms, $ V_{p}(|\bm{r}-\bm{R}|)$ shows a prominent attractive feature, the so-called butterfly potential \cite{Hamilton2002,Chibisov2002,Niederprum2016}, see Fig.\,\ref{fig:potential_energy_landscape}. For our initial states, there is a finite probability of adiabatically following the butterfly PEC, which accelerates the ground state atom towards the ionic core. At shorter internuclear distances, $V_{\text{c,g}}(R)$ takes over and further accelerates the collision process (blue shaded area in Fig.\,\ref{fig:potential_energy_landscape}). Up to this point, the dynamics can be considered as understood. For even shorter internuclear distances, the ionic core directly interacts with the ground state atom and the Rydberg electron becomes a spectator. As we will detail later on, it is this short-range physics, which is mainly responsible for two distinct processes. The first one is associative ionization, which reads for the case of rubidium

\begin{equation}
	\text{Rb}^* + \text{Rb} \rightarrow \text{Rb}_2^+ + \text{e}^- + \Delta E_{\text{b}},
\end{equation}
where $\Delta E_{\text{b}}$ is the release energy due to the chemical bond of the molecular ion. The second one is a state-changing collision, resulting in an exoergic reaction 
\begin{equation}
	\text{Rb}^* (n_\text{i},l_\text{i}) + \text{Rb} \rightarrow \text{Rb}^* (n_{\text{f}} \leq n_{\text{i}}, l_{\text{f}}) + \text{Rb} + \Delta E,
\end{equation}
where the indices i and f denote the initial or final state, respectively. $\Delta E$ is the released energy, which is transformed into kinetic energy of the ground state atom and the Ryd\-berg atom, giving rise to a dissociation of the molecule.

In the present work, we address this second type of collisions by directly measuring the momenta of the Ryd\-berg atoms after the dissociation using high-resolution state-resolved momentum spectroscopy. This method enables a clear identification of the final states (sketched as red arrows in Fig.\,\ref{fig:potential_energy_landscape}) and makes it possible to investigate the distribution of population after the collision.   

\section{\label{sectionIII}Experiment}

\begin{figure}[t]
	\includegraphics[width=\columnwidth]{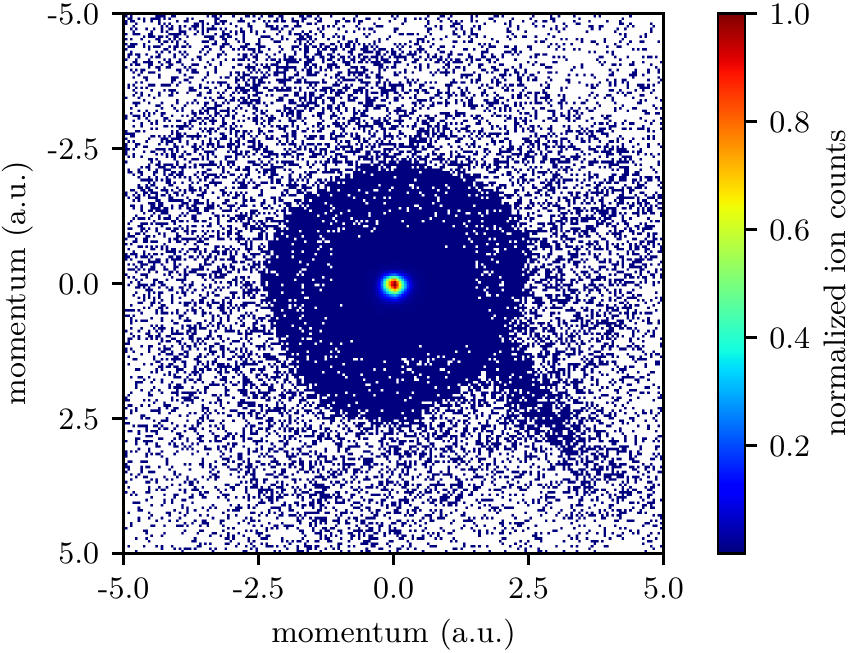}
	\caption{\label{fig:25P_momentum} Detector image of the two-dimensional projection of the three dimensional momentum distribution resulting from the decay of $25P$ ULRMs. The center of the plot consists of ions with vanishing transverse momentum, stemming mainly from photoionized ULRMs that have not undergone a state-changing collision. The two concentric circles at momenta 2.5\,a.u. and 5\,a.u. are due to  the decay into the 22Hy and and 21Hy hydrogenic manifolds. The feature in the lower right quadrant is due to technical issues of the detection unit. Sum of $10^7$ experimental runs, normalized to the maximum number of ion counts.}
\end{figure}

To measure the momentum distributions of Rydberg atoms with high resolution, we have adapted the MOTRIMS technique \cite{Wolf2000,Poel2001,Turkstra2001,Flechard2001,Nguyen2004,Blieck2008,Depaola2008,Schuricke2011,Fischer2012,Goetz2012,Hubele2015,Li2019}, included an optical dipole trap and implemented a so-called reaction microscope. The image at the top of Fig.\,\ref{fig:spectrometer} shows a three quarter section CAD drawing of the essential parts of our experimental apparatus.

Each experimental sequence starts with the trapping of pre-cooled $^{87}$Rb atoms in a three-dimensional magneto-optical trap (3D MOT). The atoms are then transferred to a crossed optical dipole trap with a wavelength of \SI{1064}{nm} (illustrated as red beams in Fig. \ref{fig:spectrometer}) and trapping frequencies of $\omega_{\text{x}}=2\pi \times \SI{2.8}{kHz}$, $\omega_{\text{y}}=2\pi \times \SI{1.4}{kHz}$ and $\omega_{\text{z}}=2\pi \times \SI{3.1}{kHz}$. After a short evaporation, the sample consists of more than $\SI{3}{}\times 10^4$ atoms, prepared in the $\ket{5S_{1/2},\text{F}=1}$ ground state, with a temperature of $\approx\SI{100}{\micro K}$ and a peak density of $\SI{1.9}{}\times 10^{13}\SI{}{atoms/cm^3}$.

Rydberg states $nP_{3/2}$ with principal quantum numbers $n$ between 20 and 27 are addressed via an off-resonant three-photon transition employing the $5P_{3/2}$ and $5D_{5/2}$ states as intermediate states. The radiation at \SI{780}{nm} (first step), \SI{776}{nm} (second step) and 1280-\SI{1310}{nm} (third step) is provided by frequency-stabilized diode laser systems. While two excitation lasers are applied from the same direction, the third one is counterpropagating (depicted as blue beam in Fig.\,\ref{fig:spectrometer}). The corresponding detunings amount to $\delta_{5P} = \SI{-60}{MHz}$ and $\delta_{5D} = +\SI{45}{MHz}$, respectively. As the photon energy of the dipole trap beams is sufficient to photoionize atoms in the $5D_{5/2}$ state (photoionization cross section $\gtrsim \SI{17}{Mb}$), the dipole trap is switched off prior to the $\SI{1}{\micro s}$ long excitation pulse. Subsequently, the atoms are recaptured, such that we can perform up to 100 experiments per sample without loosing too much ($\lesssim \SI{25}{\%}$) density.

At the bottom of Fig.\,\ref{fig:spectrometer} we illustrate the microscopic physical processes. Starting with the photoassociation of ULRMs, we wait a total of $\SI{2}{\micro s}$ during which the inelastic collisions take place and the Ryd\-berg atom changes its state (panels (a) and (b) of Fig. \ref{fig:spectrometer}). Since the final states are energetically lower than the initial state, the release energy is translated into kinetic energy, which is shared by the Rydberg atom and the ground state atom. Due to momentum conservation, both constituents move into opposite directions. Subsequently, the Rydberg atoms are photoionized by a short pulse from a high-power CO$_2$ laser (Fig. \ref{fig:spectrometer}(c)). 
With a photoionization cross section of tens of megabarns \cite{Markert2010,Gabbanini2006} the ionization process is very efficient. The recoil momentum caused by the photoionization is two orders of magnitude smaller than the typical momenta of the investigated processes, such that the created ion has the same momentum as the Rydberg atom. The ion then follows two sections of homogeneous electric fields and traverses a drift tube with zero electric field before hitting a position and time sensitive microchannel plate (MCP) delay-line detector (Fig. \ref{fig:spectrometer}(d)). This configuration is referred to as Wiley-McLaren spectrometer \cite{Wiley1955} that in particular provides space and time focusing of the ions i.e. ions with the same momentum hit the detector at the same position and the same time, independent of their initial position in the trap. As a result, we are able to measure momentum distributions of initially neutral atoms with resolutions better than $\SI{0.1}{\text{a.u.}}$ depending on the chosen electric fields. 

\section{\label{sectionIV}Results}
The outcome of our experiments are two-dimensional momentum distributions as shown in Fig.\,\ref{fig:25P_momentum} for the $25P$ state. A larger part of the ions accumulates at the center, where the transverse momentum is close to zero. These ions stem from photoionized ULRMs, which have not undergone a state-changing collision, or from facilitated off-resonantly excited Rydberg atoms \cite{Weber2015}.
Around the center, two concentric circular structures are visible. The circular shape arises from a projection of a three-dimensional spherical shell in momentum space onto the surface of the detector. The sharp boundaries of the circles thereby correspond to momentum vectors perpendicular to the spectrometer axis. As the initial state is well-defined, the energy differences $\Delta E$ to each of the lower-lying states and hence the momenta $p$ of the Rb$^+$ ions with the mass $m_{\text{Rb}}$ can be calculated using
\begin{equation}
	p = \frac{\hbar\sqrt{2 m_{\text{Rb}}\Delta E}}{a_0 E_{\text{H}}m_{\text{e}}},
\end{equation}
where $\hbar$ denotes the reduced Planck constant, $a_0$ the Bohr radius, $E_{\text{H}}$ the Hartree energy and $m_{\text{e}}$ the electron rest mass. This allows us to identify the different shells in the momentum spectra.

\begin{figure}[t]
	\includegraphics[width=\columnwidth]{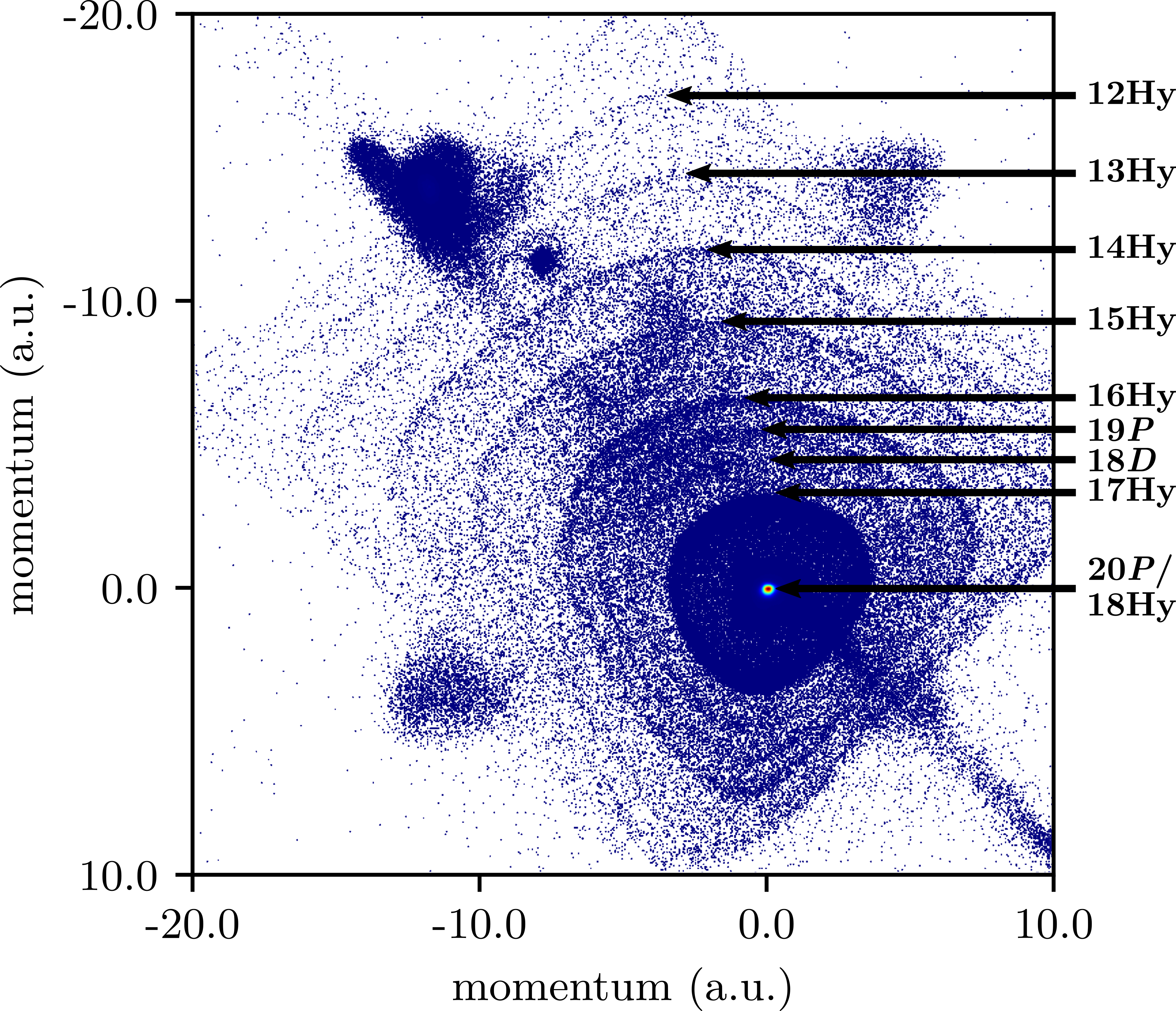}
	\caption{\label{fig:20P_momentum_full} Full detector image for state-changing collisions of 20$P$ ULRMs. Final states with principal quantum numbers down to $n=12$ are clearly visible. In addition, the decay to the $18D$- and $19P$-state can be observed, indicating the presence of low-$l$ final states. The color code is the same as in Fig. \ref{fig:25P_momentum}. Deviations from the circular shape are due to design-related imperfections, causing a repulsion of the ions along the two diagonals. The artifacts in the corners of the detector plane are due to technical issues of the detection unit.}
\end{figure}

Considering a full detector image, as shown in Fig.\,\ref{fig:20P_momentum_full} for the case of $20P$-ULRMs, a large range of final states becomes apparent. Here, we cannot only observe the decay into manifolds as low as $n=12$, also states lying in between the manifolds are clearly visible. The deviations from the circular structure are caused by design-related repulsions along the two diagonals and become more pronounced for larger momenta. Nevertheless, the final states can unambiguously be identified and evaluated.

\begin{figure}[t]
	\includegraphics[width=\columnwidth]{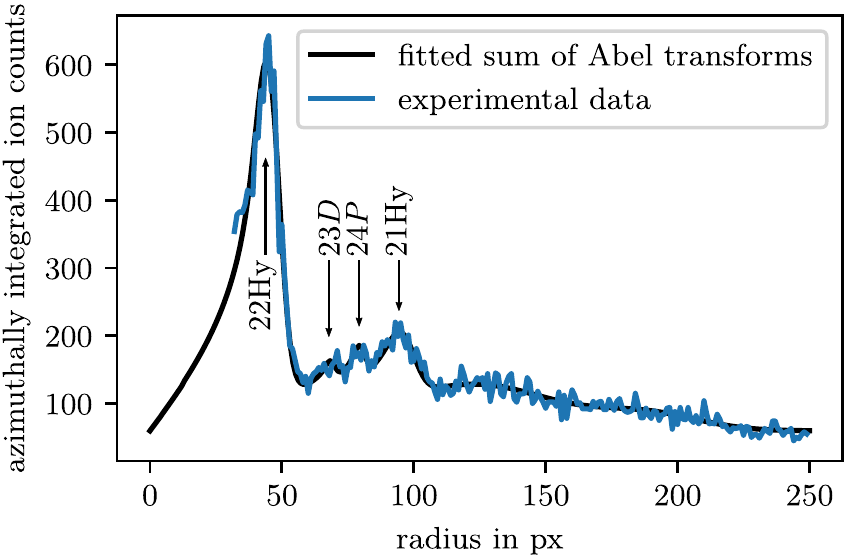}
	\caption{\label{fig:25P_radial_profile} Angular integrated radial profile of the momentum distribution shown in Fig. \ref{fig:25P_momentum} as a function of the radius. The integration is performed over a circle segment, where the spectrum shows the least distortion. The center peak is omitted due to scaling purposes. Four peaks are visible. The two most pronounced peaks correspond to manifolds 22Hy and 21Hy. In between, the $23D$ and $24P$ states are visible. We do not see signatures of a decay to the $25S$ state, which should appear between 22Hy and 23$D$. For higher radii, only the envelopes of the peaks are visible.}
\end{figure} 

We first concentrate our analysis on transitions from the initial state to the two lower-lying manifolds and the low-$l$ states in between. For a quantitative analysis of the momentum spectra we show the radial profile of the momentum distribution in Fig.\,\ref{fig:25P_radial_profile}. Since the outcome of our experiments are two-dimensional projections of initially three-dimensional spherical shells, we make use of Abel transformations. To account for the finite thickness of the shells, which corresponds to the momentum uncertainty, we assume for the three-dimensional momentum distribution a Gaussian distribution of width $\sigma$ and amplitude $A$, that is shifted isotropically by the radius $R$ of the respective shell
\begin{equation}
	f(r,R,\sigma) = \frac{A}{\sqrt{2\pi\sigma^2}} \exp\left( -\frac{(r-R)^2}{2\sigma^2}\right)
	\label{eq:Gaussian_distribution}
\end{equation}
where $r^2=x^2 + y^2 + z^2$. The two-dimensional projection along the z-axis is then given by the Abel transform of eq. (\ref{eq:Gaussian_distribution}),
\begin{eqnarray}
	F(\rho,R,\sigma) &=& \int_{-\infty}^\infty f(r,R,\sigma) \text{d}z \nonumber \\
	&=& \int_\rho^\infty 2\cdot f(r,R,\sigma) \cdot \frac{r}{\sqrt{r^2-\rho^2}}\text{d}r,
\end{eqnarray}    
with $\rho^2 = x^2 + y^2$.
            
Due to the azimuthal integration, the final fit function is given by $\rho\cdot F(\rho,R,\sigma)$. To account for the appearance of multiple peaks, we fit a sum of peaks to the data. We find good agreement with our experimental data (Fig.\,\ref{fig:25P_radial_profile}). From the fit, we extract the intensities, the momenta and their uncertainties, encoded in $A$, $R$ and $\sigma$. Particularly, this allows us to evaluate the relative amplitude for each final state. 

\begin{figure}[t]
	\includegraphics[width=\columnwidth]{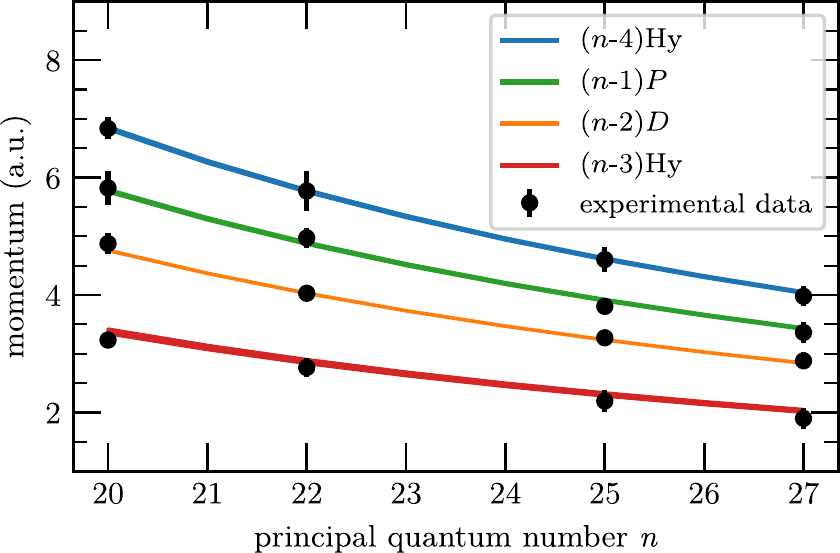}
	\caption{\label{fig:momentum_vs_n} Momenta resulting from state-changing collisions in dependence of the principal quantum number $n$ of the initial $nP$ ULRM. Depicted are the first four peaks, deduced from the fits to the respective radial profile (Fig.\,\ref{fig:25P_radial_profile}). The theoretically expected momenta are plotted as color-coded lines, where the thickness of the lines account for the momentum width (see text).}
\end{figure}

In addition to the two manifolds visible in Fig. \ref{fig:25P_momentum}, we can identify two more peaks stemming from the $23D$ and $24P$ final states. A systematic analysis for the initial quantum numbers $nP$, with $n \in \{20,22,25,27\}$ is shown in Fig. \ref{fig:momentum_vs_n}, where we plot the momenta of the fitted peaks in dependence of $n$. The expected momenta as calculated from the release energy are plotted as solid lines, where the thickness accounts for the spectral width of each final state including fine structure splitting and finite quantum defects of $F$- and $G$-states, which we count to the respective energetically close ma\-ni\-folds. In this procedure, we have introduced one global scaling factor to match experiment and theory. The perfect agreement visible in Fig.\,\ref{fig:momentum_vs_n} allows us to use this scaling factor as a calibration factor for all our momentum spectra.

In order to evaluate all detectable final states, we also include the molecular ions, which are created at short internuclear distances through associative ionization. Their signal is readily distinguished via their longer time-of-flight. The respective signals are then normalized to the total number of events $N_{\text{tot}} =  N_{\text{Rb}_2^+} + N_{\text{Rb}^+}$. The results of this first part of the analysis are summarized in Tab.\,\ref{tab:branching_ratios}.

Inspecting Tab.\,\ref{tab:branching_ratios}, several trends are noticeable. First, we note that the vast majority of events appears in the center of the momentum distribution. This is plausible as the signal in the center mainly consists of ions from long-living, photoionized ULRMs or from facilitated off-resonantly excited and photoionized Rydberg atoms. In this context, it is conspicuous that these processes appear to become less likely for initial states with lower principal quantum numbers -- in favor of state-changing collisions. Here, the decay to manifolds appears to be more likely compared with the decay to the isolated low angular momentum states in between. On the one hand, this has to do with the higher density of possible final states within a manifold. On the other hand, the autoionization resonance width of Rydberg atoms becomes narrower for larger angular momentum quantum numbers, thus suppressing the formation of molecular ions \cite{Gallagher2005}. We particularly observed a pronounced decay to the $(n-3)$ hydrogenic manifold, hence manifesting the experimental findings in \cite{Schlagmueller2016}. 

The fraction of associative ionization is in the same order of magnitude as the fraction of state-changing collisions, exhibiting no clear trend. This observation is also in close agreement with Schlagm\"uller et al. \cite{Schlagmueller2016}, who found that, starting with a 40$S$-state, the Rb$_2^+$ formation is as likely as collision-induced state-changes. 

Our results proof that the set of possible final states is not restricted to the first lower-lying hydrogenic manifold, but can in principle be extended to all lower-lying states. Also the decay into isolated low angular momentum states is possible. Since the initial state is an $nP$-state and the final state can also be a $P$-state, we call these processes state-changing collision rather than $l$-changing collisions as it is commonly referred to in literature (e.g. \cite{Higgs1981,Matsuzawa1984, Lebedev1997}).  Surprisingly, we could not find any indication of a decay into a final state with $l=0$. Evaluating our signal to noise ratio, such processes are at least suppressed by a factor of ten. These observations are new compared to previous studies \cite{Schlagmueller2016}, where only the decay into one lower-lying manifold could be experimentally observed. However, there are also  noticeable differences. In Ref.\,\cite{Schlagmueller2016}, much higher initial quantum numbers have been used. Moreover, the initial state was an $nS$-state. Looking at the molecular PECs in Fig.\,\ref{fig:potential_energy_landscape}, the initial dynamics starting from those states is quite different due to the involved quantum defects. Finally, we note that our approach is especially sensitive to small signals at high momenta, thus facilitating the detection of lower-lying states due to a high signal to noise ratio.

\begin{table}[t!]
	\caption{\label{tab:branching_ratios} Relative population of the final states down to the ($n$-4)Hy manifold and fraction of molecular ions Rb$_2^+$. All values are normalized to the sum of the detected atomic and molecular ions. The missing population distributes among lower-lying states.}
	\begin{ruledtabular}
		\begin{tabular}{ccccccc}
			\colrule
			& $nP$ & ($n$-3)Hy & ($n$-2)$D$ & ($n$-1)$P$ & ($n$-4)Hy & Rb$_2^+$\\
			\hline
			27$P$ & \SI{89.7}{\%} & \SI{1.5}{\%} & \SI{0.1}{\%} & \SI{0.01}{\%} & \SI{0.4}{\%} & \SI{3.6}{\%} \\
			25$P$ & \SI{92.1}{\%} & \SI{1.3}{\%} & \SI{0.1}{\%} & \SI{0.1}{\%} & \SI{0.4}{\%} & \SI{2.6}{\%}\\
			22$P$ & \SI{74.2}{\%} & \SI{4.0}{\%} & \SI{0.2}{\%} & \SI{0.3}{\%} & \SI{0.6}{\%} & \SI{8.2}{\%}\\
			20$P$ & \SI{69.2}{\%} & \SI{3.3}{\%} & \SI{0.3}{\%} & \SI{0.6}{\%} & \SI{1.5}{\%} & \SI{2.3}{\%} \\
		\end{tabular}
	\end{ruledtabular}        
\end{table} 

\section{\label{sectionV}Diffusive Redistribution of Population}

\begin{figure}[t]
	\includegraphics[width=\columnwidth]{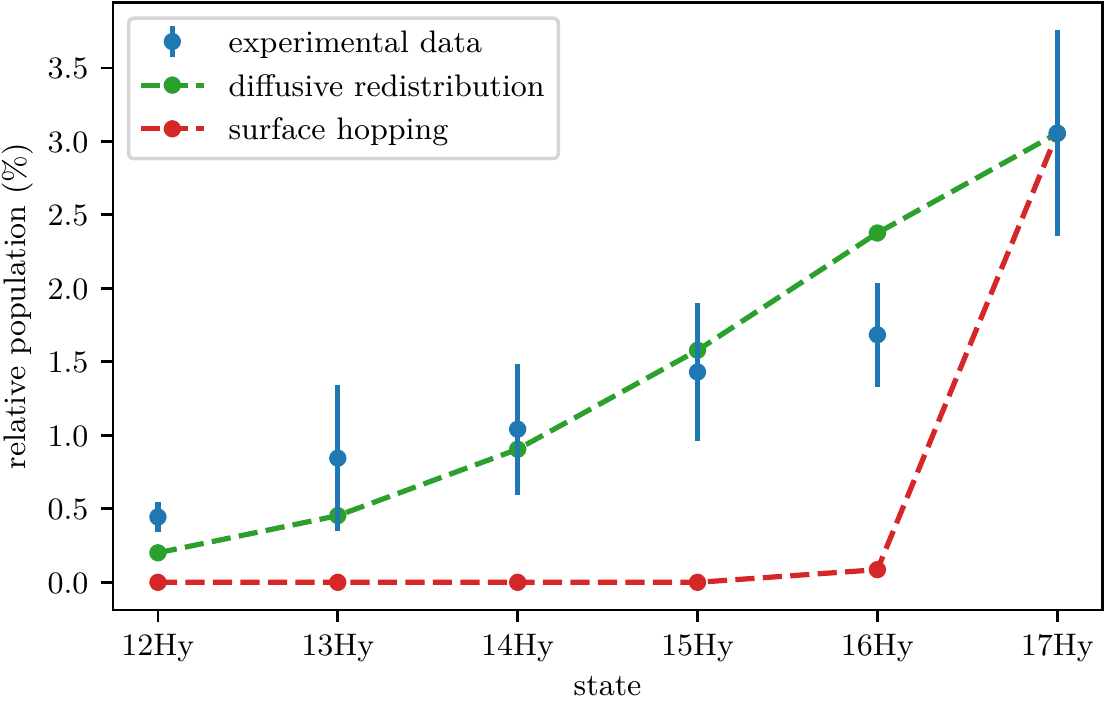}
	\caption{\label{fig:fokker_planck} Distribution of the final state population upon state-changing collisions starting from 20$P$ ULRMs. The data is normalized to the sum of events that ended up in the respective states. The diffusive model (see text) is shown as green dashed line and reproduces the experimental data well. The red dashed line shows the distribution predicted by a surface hopping model.}
\end{figure}

We now turn our attention to the details of the population distribution between the final states. In Fig.\,\ref{fig:20P_momentum_full}, one can clearly see the decay into 6 hydrogenic manifolds when starting with a 20$P$-ULRM. For simplicity, we restricted ourselves to the manifolds only and neglect low-$l$ states, which are, by comparison, substantially suppressed. The measured population in each manifold is shown in Fig.\,\ref{fig:fokker_planck}. One can clearly see a continuous decrease of the signal, with substantial weight even at the lowest detectable quantum number ($n=12$). In order to explain the wide distribution of final states, we have to look at the microscopic details of the molecular potential energy curve couplings. In Ref.\,\cite{Schlagmueller2016}, it was argued that the decay into the lower-lying manifold is due to a direct coupling of the butterfly PEC with the trilobite PEC. For an initial $nP$-state with low principal quantum number the direct coupling between the butterfly state and the trilobite state becomes small and cannot fully explain the strong decay into the next lower-lying manifold and even less the distribution among the other final states observed in our experiment. Considering couplings within the long-range part of the PEC landscape only is therefore not expedient. In fact, to understand this process one has to look in more detail at the molecular dynamics at short internuclear distances.  

\begin{figure}[t]
	\includegraphics[width=.8\columnwidth]{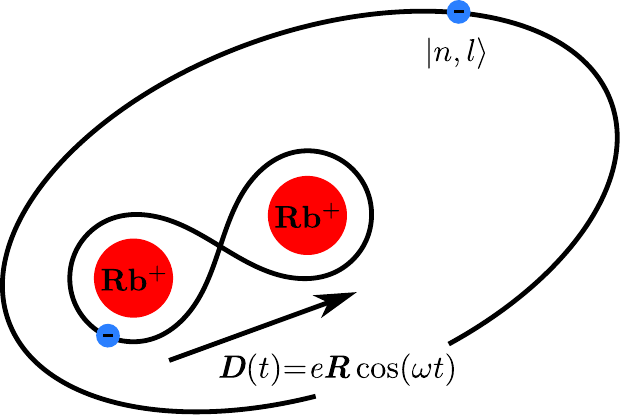}
	\caption{\label{fig:dipole_scheme} Illustration of the short-lived Rydberg quasi-molecular diatomic complex as a result of the collision of a ground state atom with a Rydberg atom. Tunneling of the inner valence electron between the two ionic cores leads to an oscillating dipole moment $\boldsymbol{D}$. The oscillating dipole induces transitions of the Rydberg electron.}
\end{figure}

Following Ref.\,\cite{Dimitrijevic2019}, inelastic collisions between Rydberg atoms and ground state atoms can often be subdivided into three phases: 1. the approach of the particles until they strongly couple to each other, 2. the formation of a Rydberg quasi-molecular complex at short internuclear distances and 3. the outcome of the collision, a state-change followed by dissociation or associative ionization. Applying this principle to the present case, the ionic core of the Rydberg atom and the neutral atom first approach each other, following the long-range PECs. When the internuclear distance has reached values $\lesssim \SI{30}{a_0}$, the subsequent dynamics can be described in the framework of a Rydberg diatomic quasi-molecular complex. This complex consists of two positively charged Rb$^+$ cores, a generalized valence electron stemming from the ground state atom and a Rydberg electron, which is shared by the molecular ionic core Rb$_2^+$ (see Fig. \ref{fig:dipole_scheme}). At such short internuclear distances, the so-called dipole resonant mechanism \cite{Smirnov1971,Mihajlov2012} becomes active. When the ionic core and the ground state atom approach each other, the valence electron starts to tunnel between the two ionic cores. This leads to an oscillating internal dipole moment $\boldsymbol{D}(t)=e\boldsymbol{R}\cos (\omega t)$ of the quasi-molecule, where $\boldsymbol{R}$ denotes the distance of the two ionic cores, $\omega$ = $\Delta(R)$ is the splitting between the gerade and ungerade wave function of the inner valence electron, and $e$ is the elementary charge. The periodic potential leads to an oscillating electric field with dipolar radiation characteristics, which can induce transitions of the Rydberg electron. In this semiclassical picture of the collisions process, the varying distance between the ionic core and the ground state atom leads to a time-varying oscillation frequency. It is therefore not surprising, that these mechanisms can induce many transitions within and between different Rydberg manifolds.

\begin{figure}[t]
	\includegraphics[width=\columnwidth]{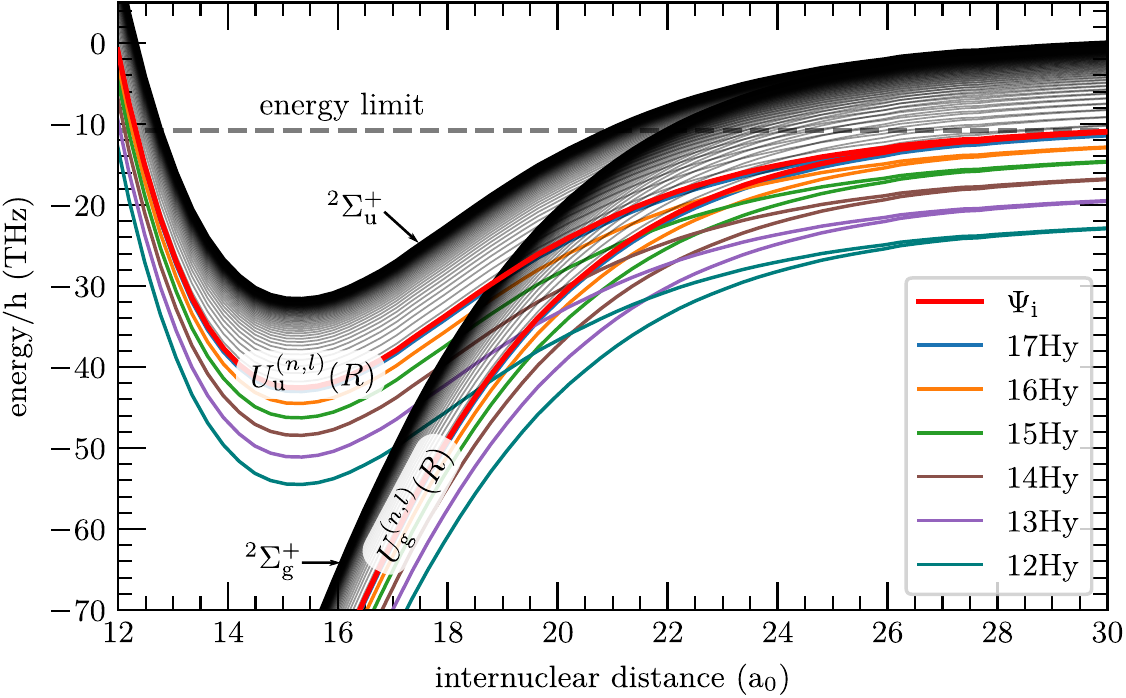}
	\caption{\label{fig:quasimolecular_potentials} Potential energy landscape of a Rydberg quasi-molecular complex. The molecular PEC split into two branches which belong the $^2\Sigma_{\text{g}}^+$ and $^2\Sigma_{\text{u}}^+$ components of the quasi-molecular ion. The highest-lying PEC of each branch belongs to $U_{\text{g}}(R)$ ($U_{\text{u}}(R)$) of the molecular ion. All other PECs $U_{\text{g,u}}^{(n,l)}(R)$ are shifted by the binding energy of the Rydberg electron. For simplicity, we restrict the plot to hydrogenic manifolds only. The multi-colored PECs are relevant for the description of the state-changing collision process when starting with 20$P$ ULRMs (see also Fig. \ref{fig:20P_momentum_full}). When the frequency of the oscillating dipole $\omega = \Delta(R)$ exceeds the binding energy of the Rydberg electron, the complex may undergo associative ionization. The gray dashed line indicates the energy limit as defined by the asymptotic potential of the initial state $\Psi_{\text{i}}$.}
\end{figure}

To illustrate these mechanisms, we show in Fig.\,\ref{fig:quasimolecular_potentials} a simplified version of the PECs at short internuclear distances. All final states of the collision when starting with 20$P$-ULRMs are highlighted. Considering only the molecular ion, Rb$_2^+$, we have two PECs, $U_{\text{g}}(R)$ and $U_{\text{u}}(R)$ for the terms $^2\Sigma_{\text{g}}^+$ and $^2\Sigma_{\text{u}}^+$. The $^2\Sigma_{\text{g}}^+$ state is the ground state of the molecular ion and forms a deep potential well. The PEC labeled as $^2\Sigma_{\text{u}}^+$, however, is predominantly repulsive with a shallow attractive section at larger internuclear distances. The energy difference between both states $\hbar \times \Delta(R)$ determines the oscillation frequency of the electron when tunneling between the two ionic cores.

The inclusion of the Rydberg electron is now done in a trivial way by only taking care of its binding energy. This results in copies of the PECs, shifted by the binding energy of the Rydberg electron,

\begin{equation}
	U_{\text{g,u}}^{(n,l)}(R) = U_{\text{g,u}}(R) + E_{\text{bind}}(n,l).
	\label{eq:quasi_molecular_potential}
\end{equation}

All other couplings, such as the spin-spin interaction, fine structure splitting, hyperfine structure effects or the exchange interaction between the two electrons, are so small that they are negligible on the energy scale given by $\hbar \times \Delta(R)$.  The relevant molecular symmetry for all PECs is therefore the gerade and ungerade one of the molecular ionic core.

\indent
The PECs in Fig.\,\ref{fig:quasimolecular_potentials} show a plenitude of crossings between the gerade and ungerade states. Due to the resonant dipole mechanism, all of these are avoided and the molecule can undergo transitions between the different PECs, resulting in an effective redistribution of the populations during the collision.

It is instructive to look at the coupling strength between the PECs. Based on the semiclassical model introduced above, we can make the following estimate: we calculated the electric field $E(t)$ induced by the oscillating internal dipole at the classical radius of the Ryd\-berg electron's orbit. We further simplify the system, by assuming that the resulting electric field is spatially homogeneous across the Rydberg electron wave function. Together with the typical transition matrix elements between neighboring Rydberg states, given by $e a_0 n^2$, we get the coupling strength $\hbar\Omega$. The resulting values of $\Omega$ are in the Terahertz range and are therefore comparable to or even exceed the energetic distance of adjacent manifolds. This complicates the molecular dynamics at short internuclear distances even further, as the coupling cannot be considered as a small perturbation to the PECs. Consequently, surface hopping models \cite{Belyaev2014} are expected to be not directly applicable. This manifests in ab initio calculations, where we calculated Landau-Zener probabilities for adiabatic and diabatic passages at each avoided crossing, thus building up a pathway through the PEC landscape. Keeping in mind that the coupling between neighboring PECs is approximately as large as the energy difference between them, it is not surprising that such a surface hopping model predicts a different behavior, see Fig.\,\ref{fig:fokker_planck}. Even though there is a redistribution, it is essentially limited to the adjacent manifolds as the transition dipole moments and therefore the couplings become increasingly smaller for direct transitions from the initial state to lower-lying states. Essentially, only the crossing with the first lower-lying manifold contributes to the redistribution process.  

To account for this strong mixing of the PECs, we adopt an effective model from Ref.\,\cite{Bezuglov2003}, where the redistribution of population between the Rydberg states is the consequence of a diffusive motion of the Rydberg electron in microwave fields \cite{Bezuglov2003}. This approach has been successfully employed for the description of collisional or thermal ionization processes \cite{Duman1980,Janev1980,Mihajlov1981,Bezuglov2001,Bezuglov2002,Bezuglov2003,Miculis2005,Mihajlov2012}. As the physical mechanisms in a state-changing collision are the same, it is applicable in our case as well. The stochastic motion of the Rydberg electron is described by a diffusion equation
\begin{equation}
	\frac{\partial}{\partial t} f(n, t) = \mathcal{D} \frac{\partial^2}{\partial n^2} f(n,t),
	\label{eq:ficks_law}
\end{equation}  
where $f(n,t)$ is the distribution of a Rydberg electron in the space of principal quantum numbers $n$ and $\mathcal{D}$ is the diffusion coefficient. Prior to the redistribution, the main principal quantum number of the initial state is given by $n_{\text{i}}$. Due to the mixing of the butterfly state we assume the population to be initially in a state with the principal quantum number $n_{\text{i}}=18$ with probability $p_i$. We then solve equation (\ref{eq:ficks_law}) using $\mathcal{D}$ as a fit parameter. The results are shown in Fig.\,\ref{fig:fokker_planck} and show good agreement with our experimental results, thus confirming a diffusive-like redistribution between the final states.

One might wonder, why the diffusive model describes the experimental observation so well, given its extreme simplicity and approximations. In fact, the exact microscopic ingredients are much more complex since the diffusion coefficient $\mathcal{D}$ in equation (\ref{eq:ficks_law}) is not necessarily a constant \cite{Dimitrijevic2019} and the oscillating electric dipole radiation field is far from being homogeneous. Moreover, we have completely ignored the $n-1$ angular momentum states of each manifold, which all have different matrix elements, and that the initial butterfly state is made up by a large number of angular momentum states.

However, when so many different initial angular momentum states and so many couplings with different strength between a plenitude of PECs contribute to the system dynamics, a diffusive-like behaviour might be, after all, just the most likely one. Our results might therefore be interpreted as a manifestation of the central limit theorem.
  
\section{\label{sectionVI}Conclusion}

We have shown that inelastic collisions between Ryd\-berg and ground state atoms can result in a large range of final states. We give evidence for the decay into low angular momentum states over a large range of principal quantum numbers. We also find pronounced decay into many lower-lying hydrogenic manifolds with substantial weight. The distribution among the manifolds suggests a diffusive-like redistribution between the Ryd\-berg states at short internuclear distances. We give a simplified explanation of this behavior in terms of redistribution of Rydberg states in microwave fields. An ab initio quantum-chemical treatment of the total collision process is a challenging task, given the different interaction mechanisms at short and large internuclear distances. Nevertheless, our results help to model parts of the collisions more accurately. Our results also have implications for the modeling of inelastic processes in many-body Rydberg systems. In the future, it will be interesting to look for effects of alignment in the initial state, where the resulting momentum distribution after the collisions becomes anisotropic. This would exploit the full 3D imaging capability of our momentum spectrometer. We expect that this development will also allow for the study of other dynamical processes in Rydberg systems such as Rydberg-Rydberg dynamics, the direct measurement of the momentum distribution of Rydberg molecules and the study of other exotic Rydberg matter, such as heavy Rydberg systems \cite{Hummel2020}.

\begin{acknowledgments}
The authors thank the group of Reinhard D\"orner (Goethe University of Frankfurt) for helpful discussions regarding the design of the MOTRIMS apparatus and Dominik Arnold for constructing the spectrometer. We thank Cihan Sahin for his help on setting up the experimental apparatus. We also thank Bergmann Messger\"ate Entwicklung KG for the excellent Pockels cell driver and the invaluable technical support. 

We acknowledge fruitful discussions with Peter Schmelcher and Fre\-de\-ric Hummel (University of Hamburg).

We gratefully acknowledge financial support by the German Research Foundation (Deutsche Forschungsgemeinschaft) within the priority programme ‘Giant Interactions in Rydberg Systems’ (DFG SPP 1929 GiRyd, project no. 316211972).
\end{acknowledgments}

\appendix
\section{Ultralong-range Rydberg Molecules}
\label{appendix:ULRM}
Ultralong-range Rydberg molecules are bound states between a Rydberg atom and at least one ground state atom. The binding results from low-energy scattering between the Rydberg electron and the ground state atom and can be expressed in the formalism of a Fermi pseudopotential \cite{Fermi1934,Omont1977,Greene2000,Hamilton2002}
\begin{eqnarray}
	V_{\text{e,g}}(|\bm{r}-\bm{R}|) &=& V_{s}(|\bm{r}-\bm{R}|) + V_{p}(|\bm{r}-\bm{R}|) \nonumber\\
	&=&2\pi a_s[k(R)] \delta(\bm{r} - \bm{R}) \nonumber\\
	&+&6\pi a_p[k(R)]\overset{\small\leftarrow}{\nabla} \delta(\bm{r}-\bm{R}) \overset{\small\rightarrow}{\nabla},	
	\label{eq:pseudopotential}
\end{eqnarray}
where $\bm{r}$ is the position of the Rydberg electron and $\bm{R}$ the position of the ground state atom with respect to the Rydberg ionic core. The first term describes $s$-wave interactions, which dominate at sufficiently large internuclear distances. At smaller internuclear distances of a few hundred Bohr radii, the $p$-wave scattering interaction comes into play. In case of alkali atoms, it equips the potential energy landscape with an attractive potential, associated with the so-called butterfly PECs \cite{Hamilton2002,Chibisov2002,Niederprum2016}, which arise from an underlying $p$-wave shape resonance. 

Besides the electron-atom scattering, one also has to account for the attractive long-range interaction between the ionic core and the polarizable ground state atom, which is given by  
\begin{equation}
	V_{\text{c,g}}(R) = -\alpha / (2R^4)
	\label{eq:ion-neutral-potential}
\end{equation}
with the polarizability $\alpha$. 

The effective Hamiltonian for the Rydberg electron therefore reads
\begin{equation}
	\mathcal{H} = \mathcal{H}_0(r) + V_{\text{c,g}}(R) + V_{\text{e,g}}(|\bm{r}-\bm{R}|),
	\label{eq:hamiltonian}
\end{equation} 
where $\mathcal{H}_0(r)$ is the Hamiltonian of the bare Rydberg atom. By diagonalizing this Hamiltonian in a finite set of basis states Born-Oppenheimer PECs can be deduced. Since the energy shift due to $V_{s}(|\bm{r}-\bm{R}|)$ is proportional to the electron probability density at the position of the perturber, the PECs are oscillatory functions of R with localized wells at extremely large separations, which can support closely spaced bound vibrational states.

\bibliography{paper_state_changing_collisions}

\end{document}